\documentclass[preprints,article,accept,moreauthors,pdftex]{Definitions/mdpi} 

\firstpage{1} 
\pubvolume{1}
\issuenum{1}
\articlenumber{0}
\pubyear{2022}
\copyrightyear{2020}
\datereceived{} 
\dateaccepted{} 
\datepublished{} 
\hreflink{https://doi.org/} 
\pdfoutput=1

 \usepackage{comment}
 
 \newcommand{\pd}{\partial}
 \newcommand{\x}{{\mathbf x}}
 
 \newcommand{\uu}{{\mathbf{u}}}

 \newcommand{\ra}{\rightarrow}

 \usepackage{placeins}

\Title{Volume Transport by a 3D Quasigeostrophic Heton}

\TitleCitation{Title}

\Author{Adhithiya Sivakumar $^{1,\ddagger}$*\orcidA{} and Jeffrey B. Weiss $^{2,\ddagger}$*\orcidB{}}

\AuthorNames{Adhithiya Sivakumar and Jeffrey B. Weiss}

\AuthorCitation{Sivakumar, A.; Weiss, J.B.}

\address{%
$^{1}$ \quad Department of Mechanical Engineering, University of New Hampshire, Durham, NH 03824\\
$^{2}$ \quad Department of Atmospheric and Oceanic Sciences, University of Colorado Boulder, Boulder, CO 80309}

\corres{Correspondence: adhithiya.sivakumar@unh.edu, jeffrey.weiss@colorado.edu}

\secondnote{These authors contributed equally to this work.}

\abstract{Oceanic flows self-organize into coherent vortices which strongly influence their transport and mixing properties. Counter-rotating vortex pairs can travel long distances and carry trapped fluid as they move. These structures are often modeled as hetons, viz. counter-rotating quasigeostrophic point vortex pairs with equal circulations. Here, we investigate the structure of the transport induced by a single three-dimensional heton. The transport is determined by the Hamiltonian structure of the velocity field induced by the heton's component vortices. The dynamics displays a sequence of bifurcations as one moves through the heton-induced velocity field in height. These bifurcations create and destroy unstable fixed points whose associated invariant manifolds bound the trapped volume. Heton configurations fall into three categories. Vertically aligned hetons do not move and do not transport fluid. Horizontally aligned hetons have a single parameter, the horizontal vortex half-separation $Y$, and simple scaling shows the dimensional trapped volume scales as $Y^3$. Tilted hetons are described by two parameters, $Y$ and the vertical vortex half-separation $Z$, rendering the scaling analysis more complex. A scaling theory is developed for the trapped volume of tilted hetons showing that it scales as $Z^4/Y$ for large $Z$. Numerical calculations illustrate the structure of the trapped volume and verify the scaling theory.}


\begin{document}

 
\section{Introduction}
Mesoscale coherent vortices are ubiquitious in the ocean and play important roles in the dynamics and transport of the ocean and climate \cite{gryanik_theory_2000, carton_hydrodynamic_2001, chelton_global_2007, chelton_global_2011, zhang_oceanic_2014, koshel_vortex_2019}. Idealized modeling of coherent vortices is a powerful tool in understanding their behavior and impact. The simplest model treats the vorticity of each coherent vortex as being concentrated at a point, resulting in a collection of point vortices. Point vortices have a long history, beginning with their application in two-dimensional fluid dynamics \cite{helmholtz_lxiii_1867, aref_point_2007}. Oceanographic flows are often approximated by the quasigeostrophic (QG) approximation where the dynamics is three-dimensional and governed by the advection of  quasigeostrophic potential vorticity (QGPV). Coherent vortices then take the form of compact regions of QGPV, which can be approximated as QG point vortices \cite{morikawa_geostrophic_1960, jg_numerical_1963, gryanik_dynamics_1983-1}.

One particularly interesting configuration of QG point vortices is the heton: a pair of counter-rotating point vortices with equal circulation magnitudes \cite{gryanik_dynamics_1983-1, hogg_heton_1985, young_interactions_1985, gryanik_theory_2000, gryanik_dynamics_2006}. Like their two-dimensional (2D) counterpart, a heton travels in a straight line, but, unlike tin 2D, the component vortices in a QG heton can be at different fluid heights. Coherent vortices in turbulent flow typically have a region of fluid trapped around them, sometimes called circulation cells, where the rotational velocity around the vortices dominates over the streaming motion further away from the vortices \cite{gryanik_theory_2000, petersen_vortex_2006}. Similarly, as a heton travels, it carries with it a volume of trapped fluid. As a result, it is able to transport the material properties of the trapped fluid over long distances. Hetons therefore often figure in models of oceanic transport and mixing processes \cite{gryanik_theory_2000, reinaud_interaction_2016, sokolovskiy_n_2020}.

Baroclinic point vortex hetons were first described by Gryanik in both the two-layer \cite{gryanik_dynamics_1983} and continuously stratified \cite{gryanik_dynamics_1983-1} formulations of QG dynamics. The term `heton' was introduced by Hogg and Stommel \cite{hogg_heton_1985}, referring to a configuration of point vortices in two-layer QG flow, with oppositely signed vortices in different layers,  producing a heat flux in the direction of movement. Studies on the trapping of passive particles by a single two-layer heton were presented in Young \cite{young_interactions_1985}, where the structure of the trapping region was found to depend on the initial vortex separation, with two distinct regimes depending on the ratio of the horizontal separation to the interfacial deformation radius.

A number of generalizations of two-layer hetons exist -- a comprehensive review of these is provided in Gryanik et al. \cite{gryanik_dynamics_2006}. Here we
study hetons in continuously stratified, unbounded, three dimensional (3D) QG flow on an $f$-plane with constant Brunt-V\"ais\"al\"a frequency and no background flow \cite{gryanik_dynamics_1983-1, gryanik_advective_1990, gryanik_interaction_1997, gryanik_theory_2000, gryanik_dynamics_2006}. Our focus is elucidating the structure of the manifolds governing the volume trapped by a single 3D heton, and investigating the magnitude of the trapped volume using both a scaling theory and numerical simulations. 

In the next section we review the equations of motion governing 3D QG hetons and passive particles in the velocity field induced by a single heton. Section 3 investigates the manifold structure governing transport and the bifurcations that give rise to the manifolds. Section 4 presents a scaling theory for the volume trapped by a heton and shows that the theory matches numerical simulations, and Section 5 contains a discussion of the results.

\section{Equations of Motion}

In this section, we review the equations governing the motion induced in a three dimensional space, $\x \equiv (x, y, z)$, by a single quasigeostrophic heton in a co-moving frame. The development, which largely follows \cite{gryanik_theory_2000}, is done in the context of a  continuously stratified fluid in an infinite domain with zero background flow and on the $f$-plane (i.e., with a constant Coriolis frequency). 

\subsection{Point Vortex Solutions to the Quasi-geostrophic Potential Vorticity Equation}
Under the situation described above, the QGPV equation and the associated streamfunction-vorticity relation are written \cite{vallis_atmospheric_2006}:
\begin{align}
    \pd_tq + \pd_x \psi \pd_y q - \pd_y \psi \pd_x q &= 0, \label{qgpv_dim}\\ 
    \pd_x^2 \psi + \pd_y^2 \psi + \pd_z \left( \frac{f^2}{N^2} \pd_z \psi\right) &= q, \label{svort_dim}
\end{align}
where $q(\x,t)$ is the quasi-geostrophic potential vorticity, $\psi$ is a streamfunction defined such that the three-dimensional velocity $\uu = (u, v, 0) = - \nabla \times \psi \hat{\mathbf{e}}_z$, $f$ is the Coriolis frequency,  $N \equiv \sqrt{-\frac{g}{\rho_0} \pd_z \rho_0}$ is the Brunt-V\"{a}is\"{a}l\"{a} frequency, which we take to be constant, and $\nabla$ is the three-dimensional gradient operator. One notable feature of QG dynamics is the the velocity is purely horizontal. Moving to stretched vertical coordinates, $z \rightarrow Nz/f$, leaves (\ref{qgpv_dim}) unchanged and replaces (\ref{svort_dim}) with a Poisson equation for $\psi$:   
\begin{align}
    \pd_tq + \pd_x \psi \pd_y q - \pd_y \psi \pd_x q &= 0, \label{qgpv_ndim}\\ 
    \nabla^2 \psi  &= q.
    \label{svort_ndim}
\end{align}
We now seek point vortex solutions of the form 
\begin{align}
    q = \sum_{i = 1}^{N_p} \Gamma_i(t) \delta(\x - \x_i(t)), \label{pv-ansat}
\end{align}
where $N_p$ is the number of point vortices, $\Gamma_i$ is the strength (or the circulation) of each point vortex, and $\x_i(t)$ its instantaneous position. Substituting (\ref{pv-ansat}) into (\ref{qgpv_ndim}), equating terms with delta functions on both sides as well as those without, we arrive at the following relations \cite{gryanik_dynamics_1983-1} \cite{gryanik_theory_2000}: 
\begin{align}
    \frac{d x_i}{dt} &= - \left.\frac{\pd \psi}{\pd y} \right|_{\x = \x_i(t)}, \label{pv-x}\\ 
    \frac{d y_i}{dt} &= \left.\frac{\pd \psi}{\pd x} \right|_{\x = \x_i(t)}, \label{pv-y}\\
    \frac{d z_i}{dt} &= 0, \nonumber\\
    \frac{d \Gamma_i}{dt} &= 0, \nonumber
\end{align}
which reveal that QG point vortices have time-invariant circulations and are constrained to travel in horizontal planes with no vertical velocity. To characterize their planar trajectories, we solve the Poisson equation ($\ref{svort_ndim}$) for $\psi(x,y,z)$. The fundamental solution associated with point sources (\ref{pv-ansat}) located instantaneously at $\x_j(t)$ is given by
\begin{align}
    \psi(\x,t) = \frac{1}{4 \pi} \sum_{j = 1}^{N_p} \frac{\Gamma_j}{\left| \x - \x_j(t)\right|}. \label{sfunc}
\end{align}
It can be seen by letting $N_p = 1$ in (\ref{sfunc}) that a single three-dimensional point vortex induces a three-dimensional horizontal velocity field whose speed decays as the horizontal distance from the vortex divided by the cube of the 3D distance from the vortex. There is a subtlety in point vortex dynamics in that it appears that the velocity at the location of a point vortex is infinite due to the denominator in $\psi$ going to zero when $\mathbf{x}=\mathbf{x}_j$. However, this is due to incorrectly taking the point vortex limit before letting $\mathbf{x}\to\mathbf{x}_j$. For an extended coherent vortex, the self-advection velocity is zero and one should first calculate the self-advection velocity, obtaining zero, and then take the point vortex limit. The QG point vortex equations of motion are then
\begin{align}
    \frac{d x_i}{dt} &= \sum_{j = 1, j \neq i}^{N_p} \frac{\Gamma_j \left(y_j - y_i\right)}{4 \pi r^3_{ij}}, \label{pvc-x}\\ 
    \frac{d y_i}{dt} &= \sum_{j = 1, j \neq i}^{N_p} \frac{\Gamma_j \left(x_i - x_j\right)}{4 \pi r^3_{ij}}, \label{pvc-y}
\end{align}
where $r_{ij} \equiv \left|\x_i - \x_j\right|$. Thus, there is no self-advection and the motion of every point vortex is due to the velocity induced by every other point vortex. The system (\ref{pvc-x} - \ref{pvc-y}) is Hamiltonian, and by virtue of its translational and rotational symmetries, conserves linear and angular momentum \cite{gryanik_dynamics_1983-1}. 

\subsection{Heton Motion}
Here we restrict our attention to a system consisting of a single heton: two vortices with equal and opposite circulations, $\Gamma =\Gamma_1 = -\Gamma_2$. As in 2D, a QG vortex pair is the simplest configuration which exhibits non-trivial dynamics.
The motion of 2D and QG vortex pairs is in many ways qualitatively similar, keeping in mind the crucial difference that QG point vortices induce a horizontal velocity field that fills the 3D space and  is a function of the 3D vector distance from the vortex.
The dynamical equations, (\ref{pvc-x} - \ref{pvc-y}), show that the velocity of each  member of a QG point vortex pair is in the horizontal direction and perpendicular to the line connecting the vortices. The vortices therefore rotate, each in their own horizontal plane, around a vertical line through their common center of vorticity, $(x_c, y_c)$, where the center of vorticity is defined as the circulation-weighted average of vortices' horizontal coordinates, \cite{helmholtz_lxiii_1867}:
\begin{align}
   x_c &\equiv \frac{\Gamma_1 x_1 + \Gamma_2 x_2}{\Gamma_1 + \Gamma_2},\\
   y_c &\equiv \frac{\Gamma_1 y_1 + \Gamma_2 y_2}{\Gamma_1 + \Gamma_2}.
\end{align}
If two vortices have circulations of the same sign, their center of vorticity lies between them. If they have circulations of opposite signs and unequal magnitudes, it lies on the extension of the line connecting the horizontal projection of their positions. For a heton, where the circulations are equal and opposite, $\Gamma_1 = -\Gamma_2$, the center of vorticity is at infinity and the vortices propagate together in a straight line.

Both 2D and QG point vortices carry with them a neighborhood of passive scalar particles as they move. Same-sign vortex pairs co-rotate and stir the neighboring fluid but, as they have no net motion, they do not induce any long-range transport. Opposite-sign vortex pairs with $\Gamma_1 + \Gamma_2 \ne 0$ travel together along curved trajectories which circle their center of vorticity. As the vortices return to their starting point, they also do not induce long-range transport. When $\Gamma_1 = -\Gamma_2$, the vortex pair propagates to infinity and does induce long-range transport. This distinction is not sharp as when $\Gamma_1 + \Gamma_2$ is small, the vortices return to their initial position only after traveling a circle with a large radius and so there is, in some sense, long-range transport. 

The equations of motion, (\ref{pvc-x}) and (\ref{pvc-y}) with $N_p = 2$ and $\Gamma = \Gamma_1 = - \Gamma_2$ reduce to 
\begin{align}
    \frac{dx_1}{dt} &= -\Gamma \frac{y_1 - y_2}{4 \pi r_{12}^3}
    \label{crpair1} \\
    \frac{dy_1}{dt} &= \Gamma \frac{x_1 - x_2}{4 \pi r_{12}^3}, 
    \\
    \frac{dx_2}{dt} &= -\Gamma \frac{y_1 - y_2}{4 \pi r_{12}^3}, 
    \\
    \frac{dy_2}{dt} &= \Gamma \frac{x_1 - x_2}{4 \pi r_{12}^3}, 
    \label{crpair4}
\end{align}
The translational invariance of the QGPV equations leads to point vortex dynamics being a function of the vortex separation, $\Delta\mathbf{x} \equiv \mathbf{x}_1 - \mathbf{x}_2$, $\Delta r \equiv |\Delta\mathbf{x}|$. In the case of a heton the vortex separation vector is constant, 
\begin{align}
    \frac{d \Delta\mathbf{x}(t)}{dt} &= 0 
\end{align}
This is a unique result of the heton. For other vortex pairs, $\Gamma_1+\Gamma_2\ne 0$, the separation distance $\Delta r$ remains constant but the direction of $\Delta\mathbf{x}$ changes as the pair rotates. Defining the horizontal velocity
\begin{align}
\mathbf{U} &= 
\begin{pmatrix}
U\\V
\end{pmatrix}
=\frac{\Gamma}{4 \pi \Delta r^3}
\begin{pmatrix}
-\Delta y\\
\hspace{0.8em}\Delta x
\end{pmatrix},
\end{align}
which is constant in time since $\Delta\mathbf{x}$ is constant,
allows the dynamics of the component vortices of the heton, (\ref{crpair1} - \ref{crpair4}), to be written as
\begin{align}
    \frac{d\mathbf{x}_1}{dt} &= \frac{d\mathbf{x}_2}{dt} = \mathbf{U}. \label{crpair11}   
\end{align}
Thus, the members of a 3D heton travel together in a straight line with horizontal velocity $\mathbf{U}$ and, as is always the case in QG dynamics, with zero vertical velocity. Note that if the vortices are vertically aligned, $\Delta x = \Delta y = 0$, $\Delta z \ne 0$, then $\mathbf{U} = 0$ and the heton is stationary.

\subsection{The Passive Scalar Motion Induced by an Isolated 3D Heton}

\begin{figure}[H]
\centering
\includegraphics[width=0.75\linewidth]{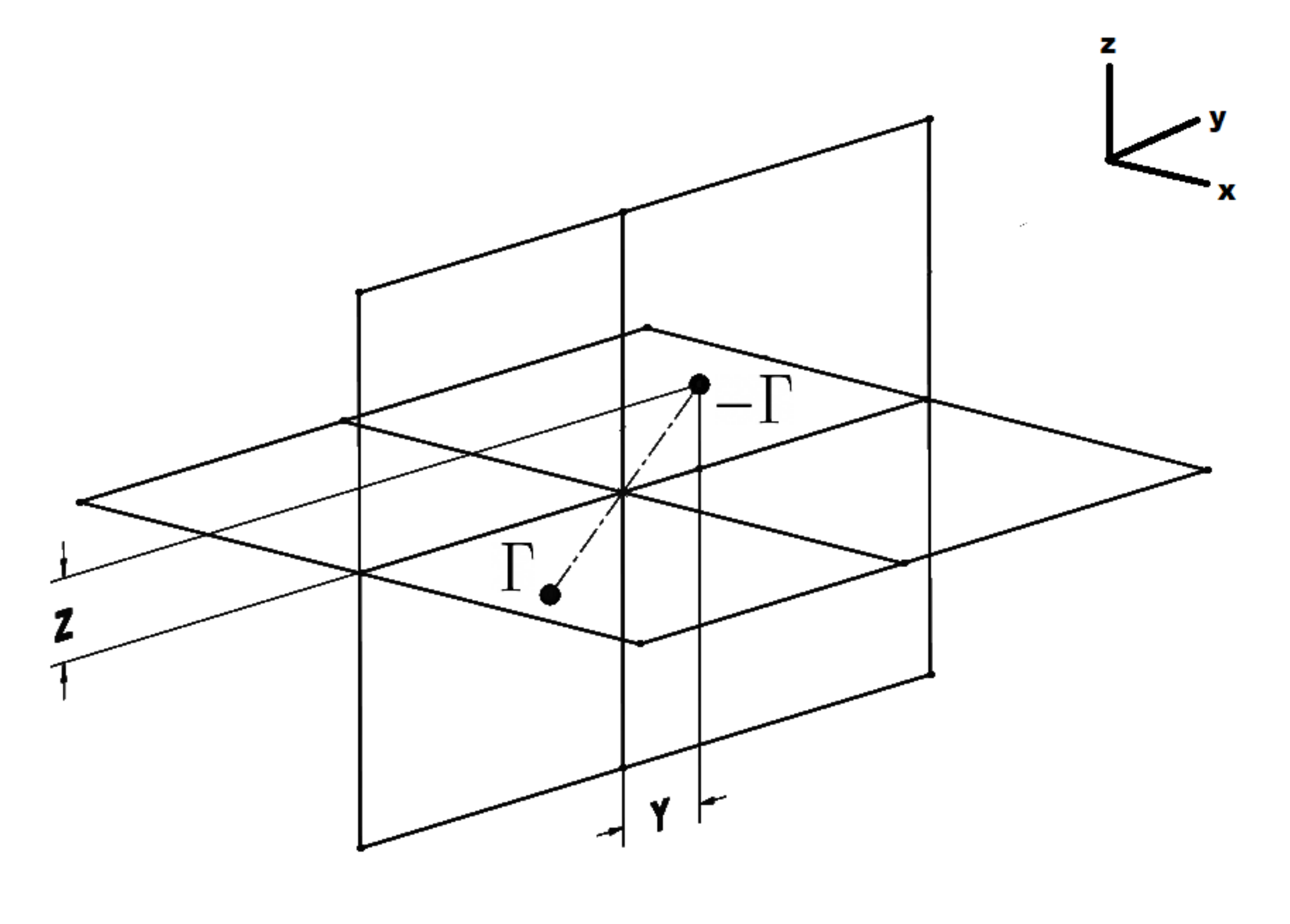}
\caption{Schematic diagram of a three-dimensional heton with an anticyclonic point vortex (circulation $-\Gamma$) initially located at $(0, Y, Z)$ and a cyclonic vortex (circulation $\Gamma$) initially located at $(0, -Y, -Z)$.}
\label{fig1}
\end{figure}

A collection of point vortices induces a velocity field that fills the space of the fluid. Passive scalar particles move with this induced velocity field. Here we consider the motion of a passive scalar located at $\x(t)$ away from the vortices, $\x(t) \ne \x_i(t)$.
The velocity field induced by a 3D QG heton is:
\begin{align}
    \frac{dx}{dt} &= \frac{\Gamma(y - y_1)}{4 \pi |\x - \x_1|^3} - \frac{\Gamma(y - y_2)}{4 \pi |\x - \x_2|^3}, \label{hetgen1} \\
    \frac{dy}{dt} &= -\frac{\Gamma(x - x_1)}{4 \pi |\x - \x_1|^3} + \frac{\Gamma(x - x_2)}{4 \pi |\x - \x_2|2^3}, \label{hetgen2} \\
    \frac{dz}{dt} &= 0. \label{hetgen3}
\end{align}
The dynamics (\ref{hetgen1} - \ref{hetgen2}) possesses symmetries of translation in the horizontal plane and rotation about a vertical axis passing through the midpoint of the line connecting the vortices. Without loss of generality, therefore, we may align the vortex initial condition on the $y$-axis, centered on the origin. Defining the vortex separation as $\Delta\x = (0, 2Y, 2Z)$, $Y, Z \geq 0$, gives the initial vortex positions as $\x_1(0) =(0, Y, Z)$ and $\x_2(0) = (0, -Y, -Z)$, as depicted in Figure \ref{fig1}. 
In this configuration, the heton undergoes translation along the $x$-axis with a velocity 
\begin{align}
    U = \frac{-\Gamma Y}{16 \pi (Y^2 + Z^2)^{3/2}}, \quad V = 0, \label{hetvel}
\end{align}
and induces a velocity field given by 
\begin{align}
    \frac{dx}{dt} &= \frac{\Gamma}{4 \pi} \left \{ \frac{y - Y}{\left[ (x - Ut)^2 + (y-Y)^2 + (z - Z)^2\right]^{3/2}} \right.\nonumber\\
    &\qquad\qquad\qquad - \left.\frac{y + Y}{\left[ (x - Ut)^2 + (y+Y)^2 + (z + Z)^2\right]^{3/2}}\right\}, \\
    \frac{dy}{dt} &= \frac{\Gamma}{4 \pi} \left \{ \frac{x - Ut}{\left[ (x-Ut)^2 + (y+Y)^2 + (z + Z)^2\right]^{3/2}} \right.\nonumber\\
    &\qquad\qquad\qquad - \left. \frac{x - Ut}{\left[ (x-Ut)^2 + (y-Y)^2 + (z - Z)^2\right]^{3/2}}\right\}. 
\end{align}
The dynamics simplifies in a frame moving with the heton, i.e., we apply the change of variable $x \ra x + Ut$, under which the above equations of motion become autonomous:
\begin{align}
    \frac{dx}{dt} &= \frac{\Gamma}{4 \pi} \left \{ \frac{y - Y}{\left[ x^2 + (y-Y)^2 + (z - Z)^2\right]^{3/2}} - \frac{y + Y}{\left[ x^2 + (y+Y)^2 + (z + Z)^2\right]^{3/2}}\right\} - U, \label{hetmovx} \\
    \frac{dy}{dt} &= \frac{\Gamma}{4 \pi} \left \{ \frac{x}{\left[ x^2 + (y+Y)^2 + (z + Z)^2\right]^{3/2}} - \frac{x}{\left[ x^2 + (y-Y)^2 + (z - Z)^2\right]^{3/2}}\right\}. \label{hetmovy} 
\end{align}
The velocity field above is associated with a streamfunction $\psi(\x)$ given by 
\begin{align}
    \psi &\equiv \frac{\Gamma}{4 \pi} \left \{ \frac{1}{\left[ x^2 + (y-Y)^2 + (z - Z)^2\right]^{1/2}} - \frac{1}{\left[ x^2 + (y+Y)^2 + (z + Z)^2\right]^{1/2}}\right\} + Uy.\\
    \frac{dx}{dt} &= -\frac{\partial \psi}{\partial y},\label{hetmstr1}
    \\
    \frac{dy}{dt} &= \frac{\partial \psi}{\partial x}
    \label{hetmstr}
\end{align}
One sees that the dynamics (\ref{hetmstr1})-(\ref{hetmstr}) is Hamiltonian where $\psi$ is the Hamiltonian and the horizontal positions of the passive scalars the canonical coordinates. This is a general feature of both 2D and QG hetons and reflects the fact that the fluid motion is 2D and incompressible. 

The above equations of motion for passive scalars in the field of a 3D heton manifestly depend on three parameters, $\Gamma$, $Y$, and $Z$. We choose units of time so that $\Gamma = 1$. One is also free to choose a unit of length. However one cannot independently choose vertical and horizontal length scales as the ratio between these length scales has already been chosen to make the Brunt-V\"{a}is\"{a}l\"{a} frequency $N = 1$. One possible scaling is to choose a length scale so that $Y=1$. This choice precludes investigating  a heton with vertically aligned vortices, i.e., $Y=0$. Alternatively, one can choose a length scale so that $Z=1$, which precludes investigating a heton with vortices on the same horizontal plane, $Z=0$. 

The dynamics of vertically aligned hetons, $Y = 0$, is relatively simple and rescaling does not provide additional insight. For horizontally offset hetons, $Y\ne 0$, we scale length and time such that $Y=1$ and $\Gamma = 1$, 
\begin{align}
\mathbf{x}'&= \mathbf{x}/Y,\qquad
t' = \Gamma t /Y^3, \qquad \psi' = Y \psi/\Gamma \label{rescale} 
\end{align} and drop primes. The resulting nondimensional equations of motion are:
\begin{align}
    \frac{dx}{dt} &= \frac{1}{4 \pi} \left \{ \frac{y - 1}{\left[ x^2 + (y-1)^2 + (z - Z)^2\right]^{3/2}} - \frac{y + 1}{\left[ x^2 + (y+1)^2 + (z + Z)^2\right]^{3/2}}\right\} - U, \label{hetmovxII} \\
    \frac{dy}{dt} &= \frac{1}{4 \pi} \left \{ \frac{x}{\left[ x^2 + (y+1)^2 + (z + Z)^2\right]^{3/2}} - \frac{x}{\left[ x^2 + (y-1)^2 + (z - Z)^2\right]^{3/2}}\right\}, \label{hetmovyII}  \\
    \psi &= \frac{1}{4 \pi} \left \{ \frac{1}{\left[ x^2 + (y-1)^2 + (z - Z)^2\right]^{1/2}} - \frac{1}{\left[ x^2 + (y+1)^2 + (z + Z)^2\right]^{1/2}}\right\} + Uy, \label{hetmstrII}\\
    U &= - \frac{1}{16\pi(1 + Z^2)^{3/2}}. \label{hetvelII}
\end{align}
The 3D dynamics is a function of a single parameter, the nondimensional vertical vortex half-separation $Z$. For each vortex configuration with fixed $Z$, the 3D dynamics foliates into independent 2D dynamical systems, parameterized by the height $z$ of the horizontal plane. Further, note that the nondimensional $Z$ obeys the relation \[Z = \frac{N}{f} \frac{Z_{dim}}{Y_{dim}},\] where the subscripts denote dimensional quantities. Typical values of $N$ and $f$ in the ocean at midlatitudes are $N = 10^{-2}$ $s^{-1}$ and $f = 10^{-4}$ $s^{-1}$ \cite{vallis_atmospheric_2006}, so $Z \sim O(1)$ corresponds to a heton with a dimensional aspect ratio of $Z_{dim}/Y_{dim} \sim O\left(10^{-2}\right)$. 

\section{The Structure of Heton-Induced Transport}

As a heton travels, it carries with it a trapped volume of fluid. Fluid within the trapped volume moves with the heton and rotates around the vortices as they move. Fluid outside this volume may get dragged along by the heton for a finite distance before ultimately being left behind. 

As described in the above section, we work in frame co-moving with the heton. In this co-moving frame, the trapped fluid has no net motion, and the untrapped fluid moves past the stationary heton with trajectories extending from infinitely far in front of the heton, where "front" refers to the direction the heton is heading, to infinitely far behind the heton. The heton is stationary in the co-moving frame and the trapped fluid rotates around the component vortices without leaving their neighborhood. In this co-moving frame, the trapped region is defined by the invariant stable and unstable manifolds associated with the equations (\ref{hetmovx} - \ref{hetmstr}). 

These manifolds by definition belong to hyperbolic fixed points of the co-moving dynamics, and, since the system is Hamiltonian, they coincide with streamlines of the flow with the same value of $\psi$ as their  associated fixed point. The structure of heton-induced transport is thus governed by the structure of the fixed-points and their associated manifolds in the co-moving frame. It is useful to separately consider the three possible heton configurations: a vertically aligned heton with $Y=0$, $Z \ne 0$ (We can scale $Z=1$ if we want); a horizontally aligned heton with $Y = 1$, $Z = 0$; and the more general tilted heton with $Y = 1$, $Z\ne 0$. 

\subsection{Vertically aligned heton: $Y = 0$, $Z \ne 0$}

A vertically aligned heton, $Y=0$, $Z\ne 0$, is stationary. The induced velocity field takes the form of axisymmetric rotation about the $z$-axis. There are no fixed points in velocity field and, thus, no manifolds. Since the heton is stationary and all trajectories rotate around the heton, the volume trapped by a vertically aligned heton is, in a sense, infinite. This has analogues in the two-layer case, as noted in \cite{young_interactions_1985} and \cite{hogg_heton_1985}.

\subsection{Horizontally aligned heton: $Y=1$, $Z = 0$}

For this case the heton equations (\ref{hetmovxII} - \ref{hetvelII}) take the form:
\begin{align}
    \frac{dx}{dt} &= \frac{1}{4 \pi} \left \{ \frac{y - 1}{\left[ x^2 + (y-1)^2 + z^2\right]^{3/2}} - \frac{y + 1}{\left[ x^2 + (y+1)^2 + z^2\right]^{3/2}}\right\} + \frac{1}{16 \pi}, \label{hetmovxI} \\
    \frac{dy}{dt} &= \frac{1}{4 \pi} \left \{ \frac{x}{\left[ x^2 + (y+1)^2 + z^2\right]^{3/2}} - \frac{x}{\left[ x^2 + (y-1)^2 + z^2\right]^{3/2}}\right\},\label{hetmovyI} \\
    \psi &= \frac{1}{4 \pi} \left \{ \frac{1}{\left[ x^2 + (y-1)^2 + z^2\right]^{1/2}} - \frac{1}{\left[ x^2 + (y+1)^2 + z^2\right]^{1/2}}\right\} - \frac{y}{16 \pi}. \label{hetmstrI}
\end{align}
Since, with $Z=0$, the equations remain unchanged under the transformation $z \ra -z$, the dynamics are symmetric about the plane $z = 0$. 

\begin{figure}[H]
\includegraphics[width=0.5\linewidth]{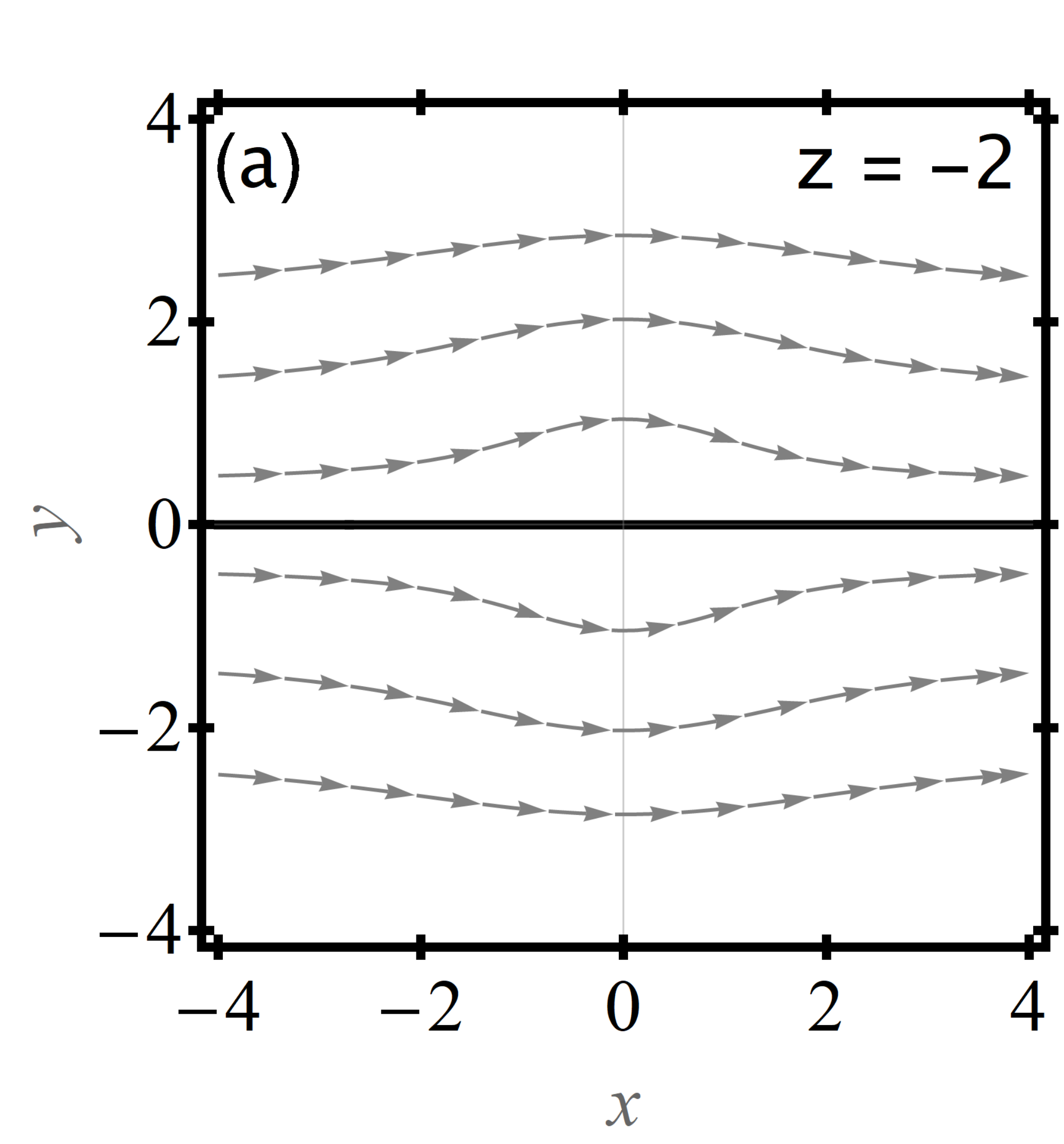}
\includegraphics[width=0.5\linewidth]{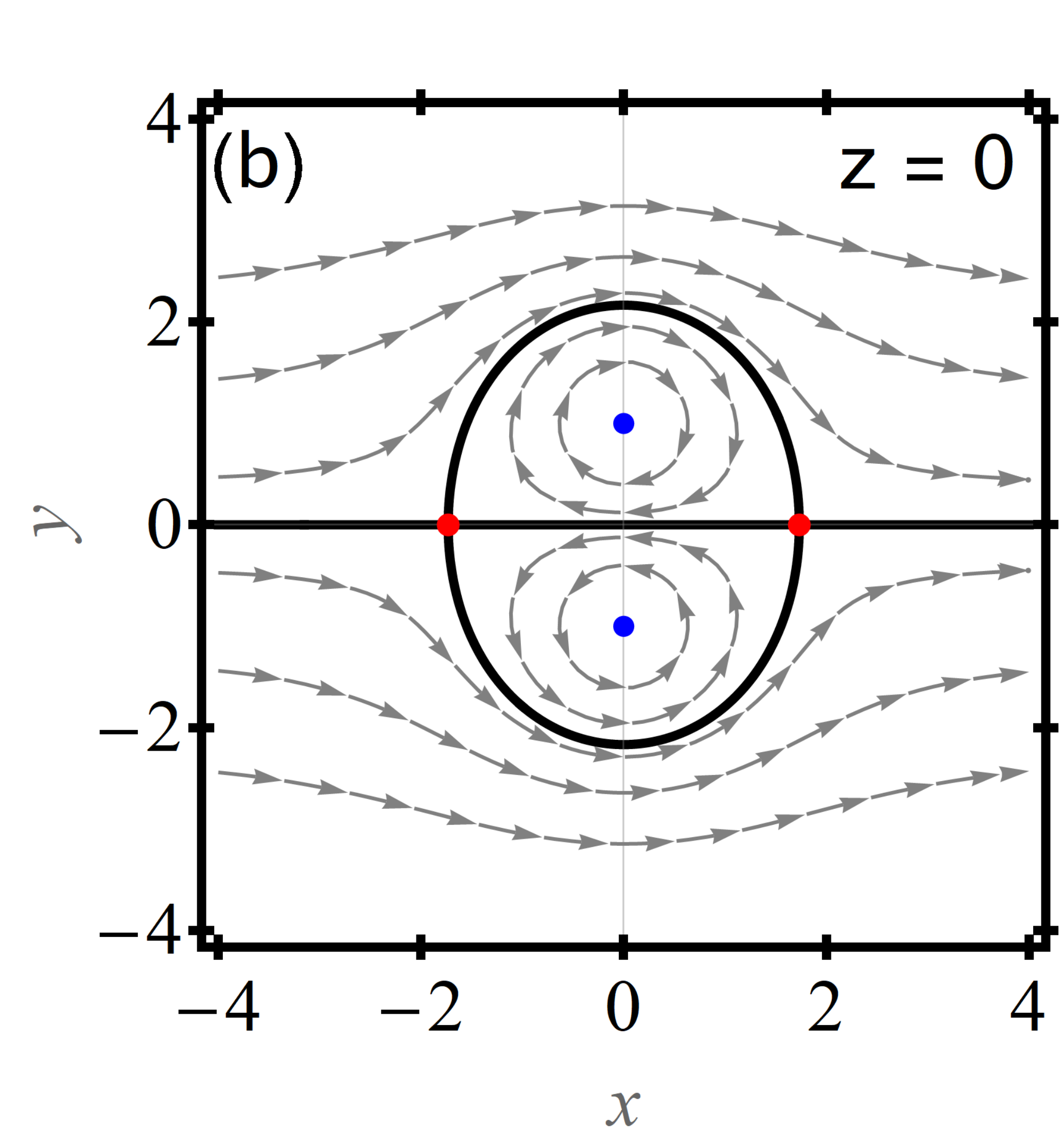}
\caption{Phase portraits of the flow induced by a heton with $Z = 0$ on horizontal planes with (a) $z = -2$, and (b) $z = 0$. These plots show the normalized planar velocity field (grey arrows), the heteroclinic separatrix $\psi = 0$ (black) which bounds the trapping region, and stable and unstable fixed points (blue and red dots, respectively). In panel (a), at `large' distances from the heton, no fluid is trapped. Closer to the heton, as in panel (b), a trapping region exists.}
\label{fig2}
\end{figure}

At heights far above and below the heton, the heton induced velocity is weak, the flow in the co-moving frame is dominated by the drift due to being in a moving frame, and there is no trapping (Figure \ref{fig2}a). On the plane containing the vortices, the flow field contains four fixed points: two unstable saddles and two stable centers (Figure \ref{fig2}b). As is typical for conservative systems, a separatrix connects the saddles and bounds the trapped region. 
\FloatBarrier

\begin{figure}[H]
\includegraphics[width=\linewidth]{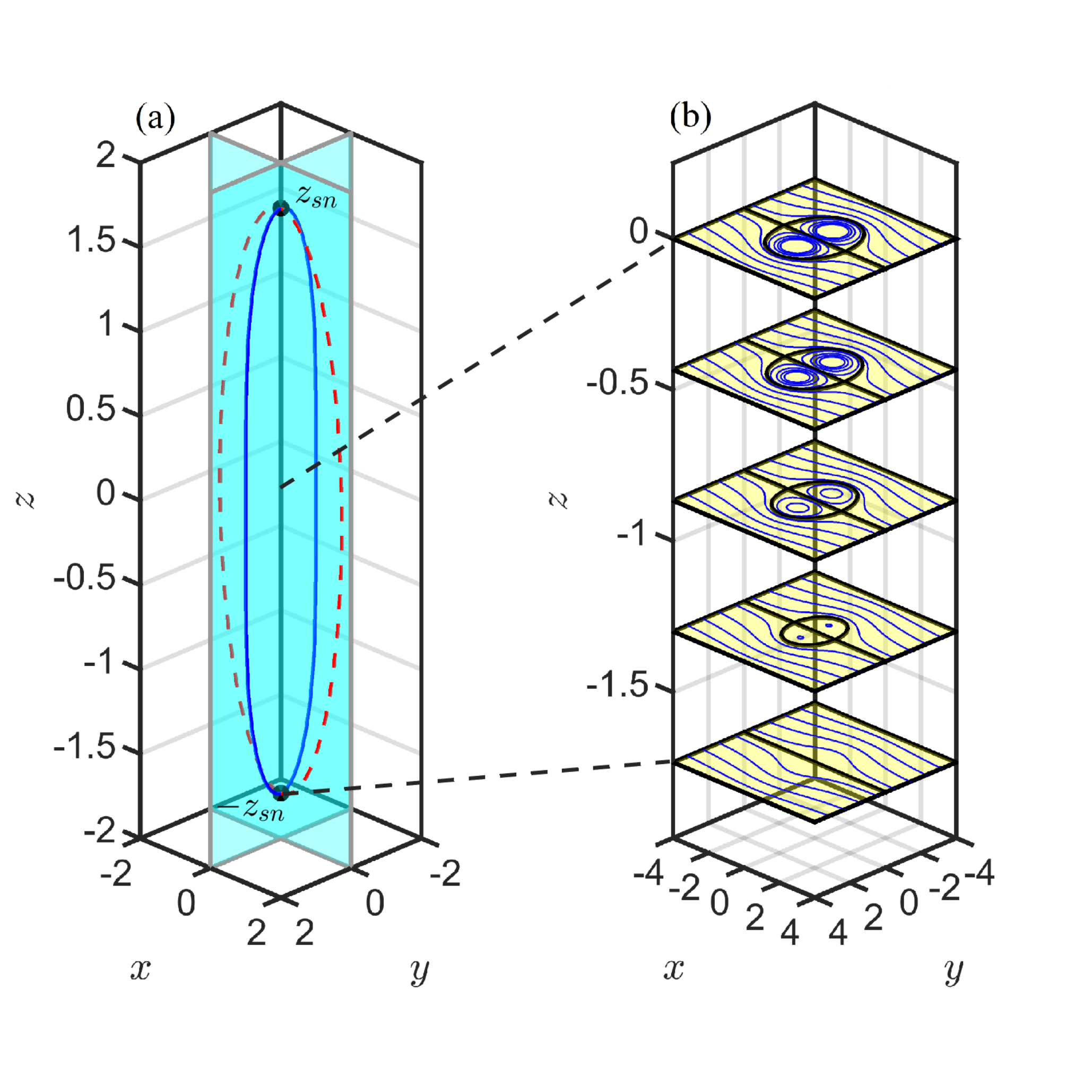}
\caption{Bifurcation diagram (a) and contour slices (b) for the $Z = 0$ case. Panel (a) illustrates the generation/destruction of stable (solid blue curves) and unstable (dashed red curves) fixed points. At $z = -z_{sn} = -\sqrt{3}$, a double saddle-node bifurcation leads to the creation of two pairs of fixed points. The stable pair is on the $x = 0$ plane, and the unstable pair is on the $y = 0$ plane. These fixed points are destroyed by another double saddle-node bifurcation at $z = z_{sn} = \sqrt{3}$. The contour slices of panel (b) illustrate the three-dimensional volume trapped by the heton at five equally spaced planes between $z = -\sqrt{3}$ and $z = 0$.}
\label{fig3}
\end{figure}

One can treat the height $z$ as a bifurcation parameter. As one moves up from $z=-\infty$, the four fixed points are created by a double saddle-node bifurcation at some $z=-z_{sn}$ and remain as one moves to $z=0$. As $z$ increases through zero and to $+\infty$, symmetry shows that the fixed points disappear at $z=+z_{sn}$ through another double saddle-node bifurcation. The bifurcation diagram is shown in Figure \ref{fig3}a. 

The location of the unstable fixed points can, for this case, be determined analytically. By symmetry, they occur at $y= 0$. Then (\ref{hetmovyI}) gives $dy/dt = 0$ for all $x$. Equating (\ref{hetmovxI}) to zero along with the conditions $x \neq 0$, $y = 0$, yields the fixed points
$\x = \left(\pm \sqrt{3 - z^2},0, z \right)$. 
The unstable fixed points are thus farthest apart in $x$ when $z = 0$, i.e., in the plane containing the vortices, and the horizontal region trapped by the heton is thus largest on this plane. As one moves away from this plane, the trapping region shrinks, eventually disappearing altogether on the planes $z=\pm z_{sn}$ at the double saddle-node bifurcations. Requiring that the $x$-coordinate of the fixed point be real shows that the saddle-node bifurcation occurs at $z_{sn}=\sqrt{3}$. The stable fixed points are determined by repeating the process above with $x = 0$. Again, $dy/dt$ vanishes. The equation for the zeroes of $dx/dt$ now, however, must be found numerically. 

\FloatBarrier
The three-dimensional volume trapped by the heton can be inferred from the contour slices seen in Figure \ref{fig3}b. The case under consideration here, $Y=1$, $Z=0$, has no remaining free parameters. The trapped volume is therefore a finite constant. Reversing the non-dimensionalization (\ref{rescale}) reveals that volume of trapped fluid scales as $Y^3$, the cube of the horizontal half-separation between the component vortices of the heton. 

\subsection{Tilted heton: $Y=1$, $Z \neq 0$}

The most general case, a tilted heton with $Y=1$ and $Z \ne 0$, is described by equations (\ref{hetmovxII} - \ref{hetvelII}). Tilting the heton breaks the symmetry in $z$ and the dynamics differs across the $z=0$ plane. However, the equations  (\ref{hetmovxII} - \ref{hetvelII}) do remain unchanged under the transformation
\begin{equation}
    z \ra -z,\,\, t \ra -t,\,\, \psi \ra -\psi.
\end{equation}
We note that a more physical version of this symmetry is to keep time moving forward but change the signs of the vortex circulations. This "mirror" symmetry is useful in simplifying the analysis. 

\begin{figure}[H]
\includegraphics[width=0.5\linewidth]{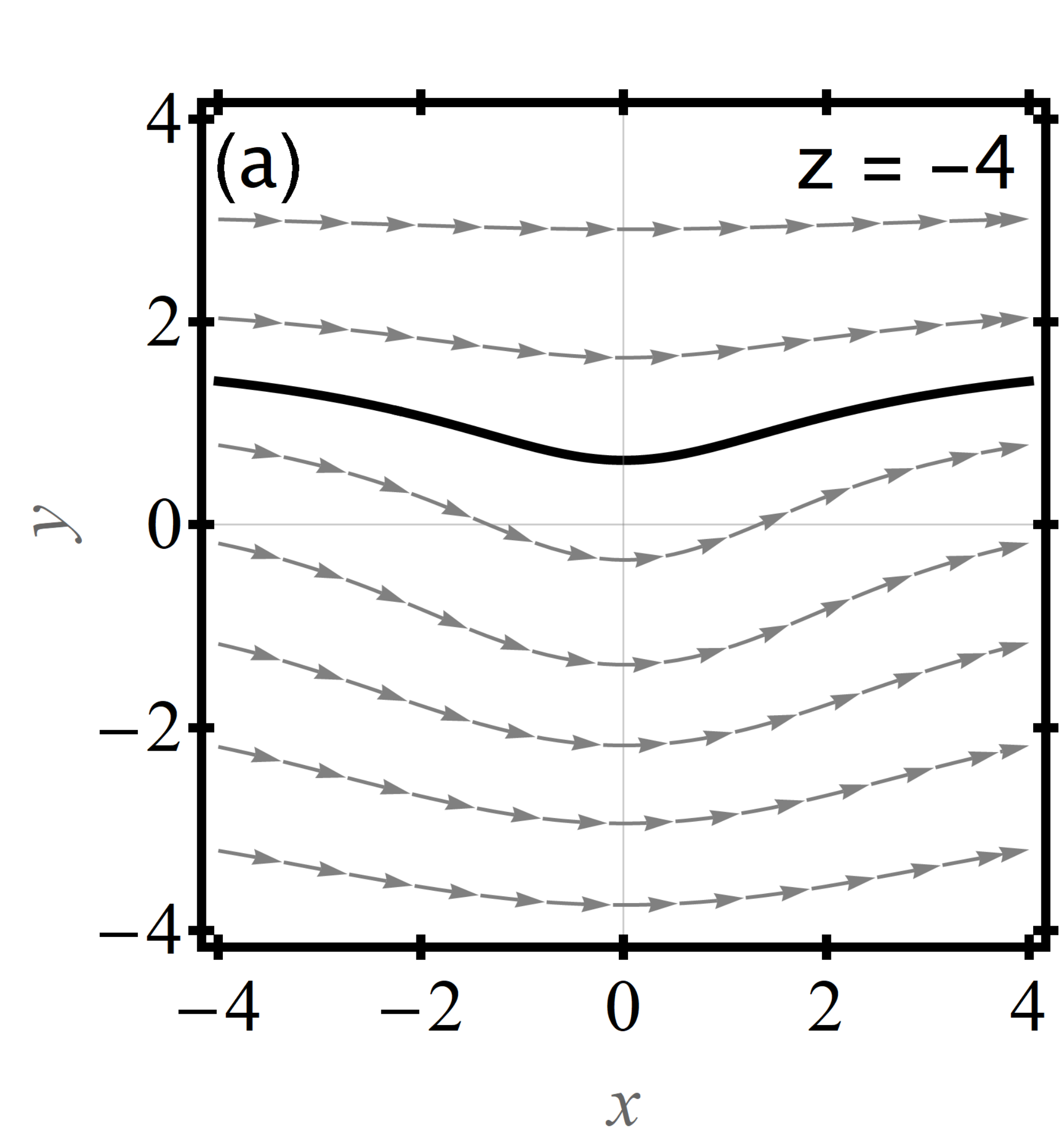}
\includegraphics[width=0.5\linewidth]{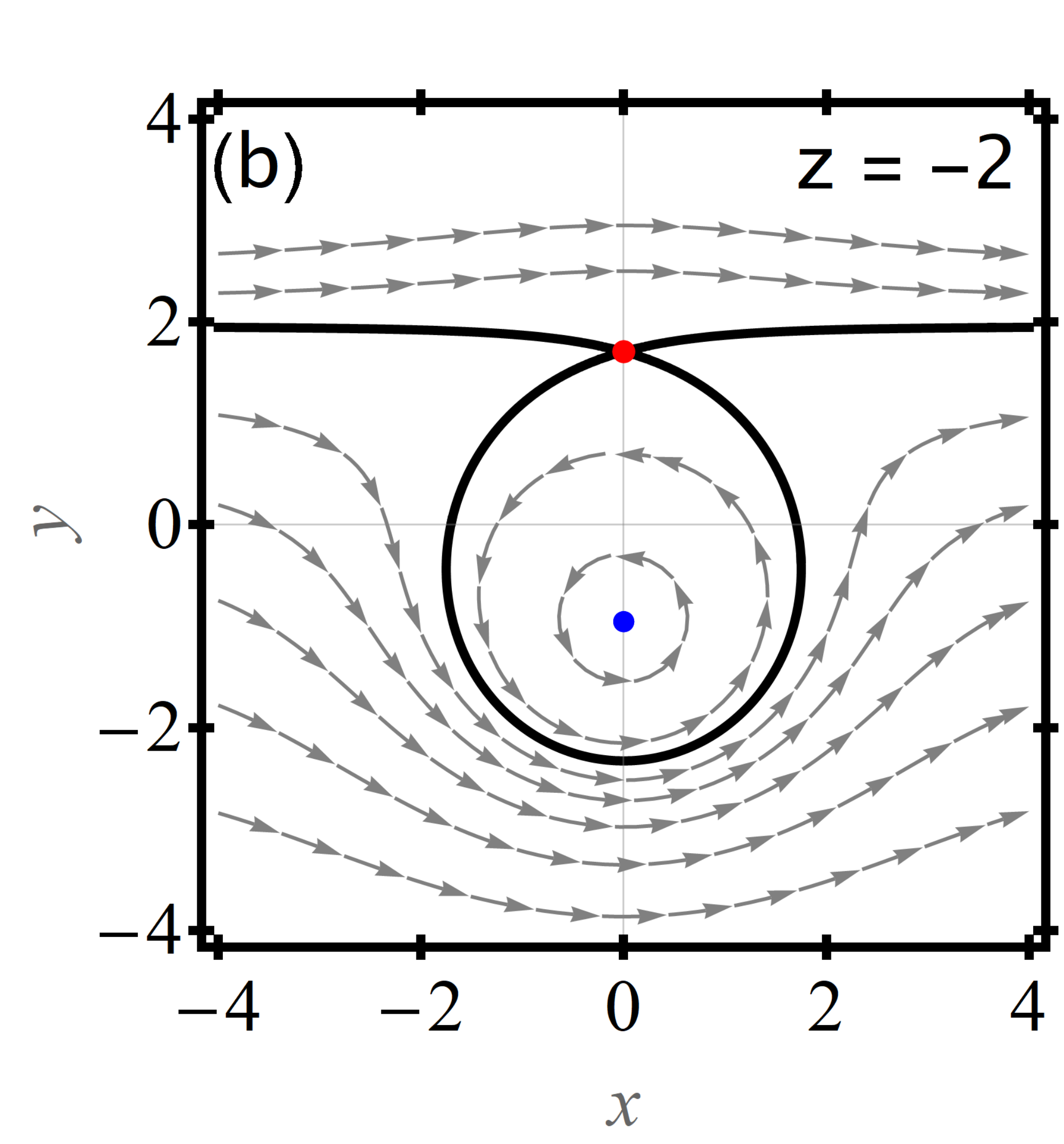}
\includegraphics[width=0.5\linewidth]{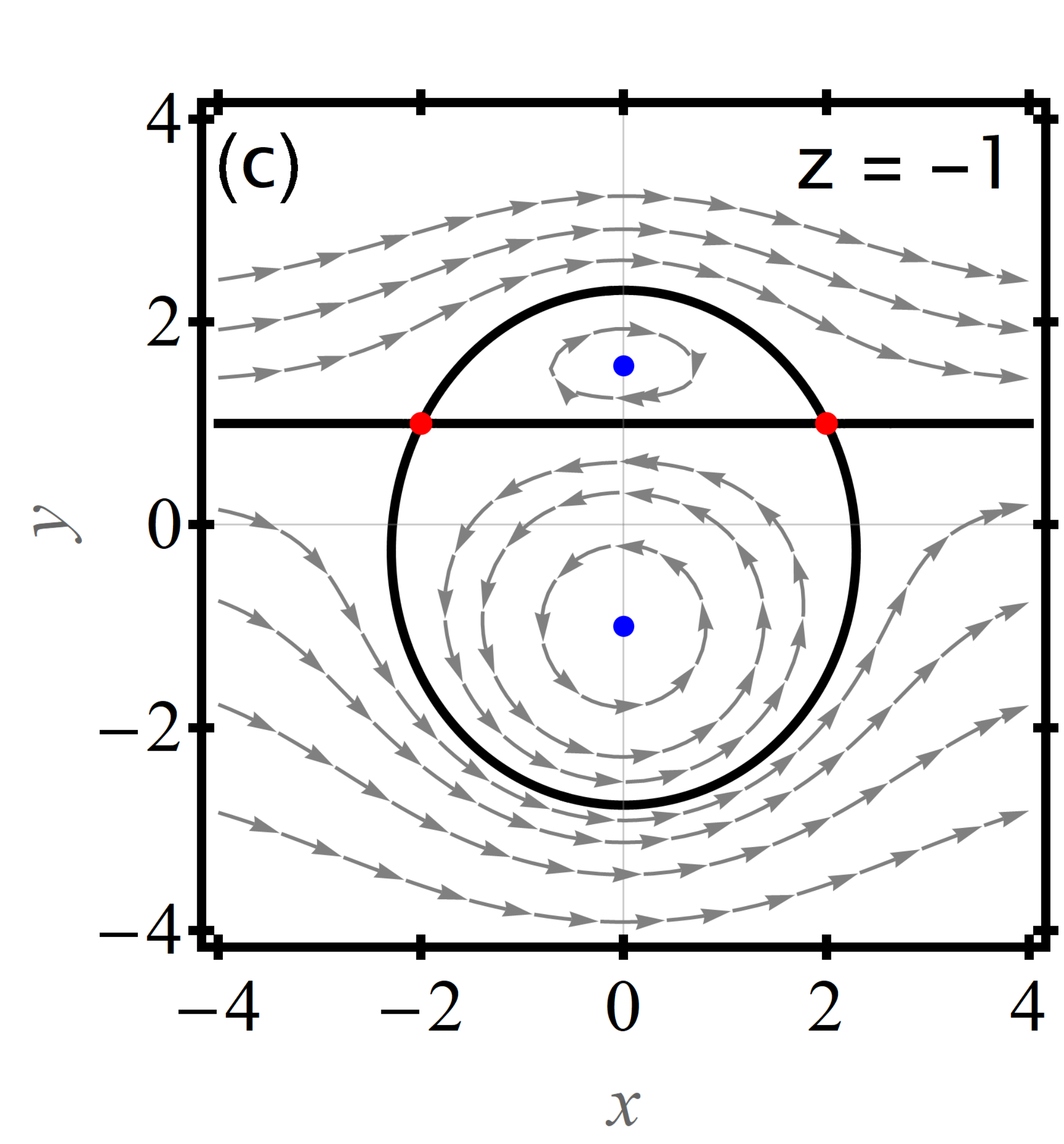}
\includegraphics[width=0.5\linewidth]{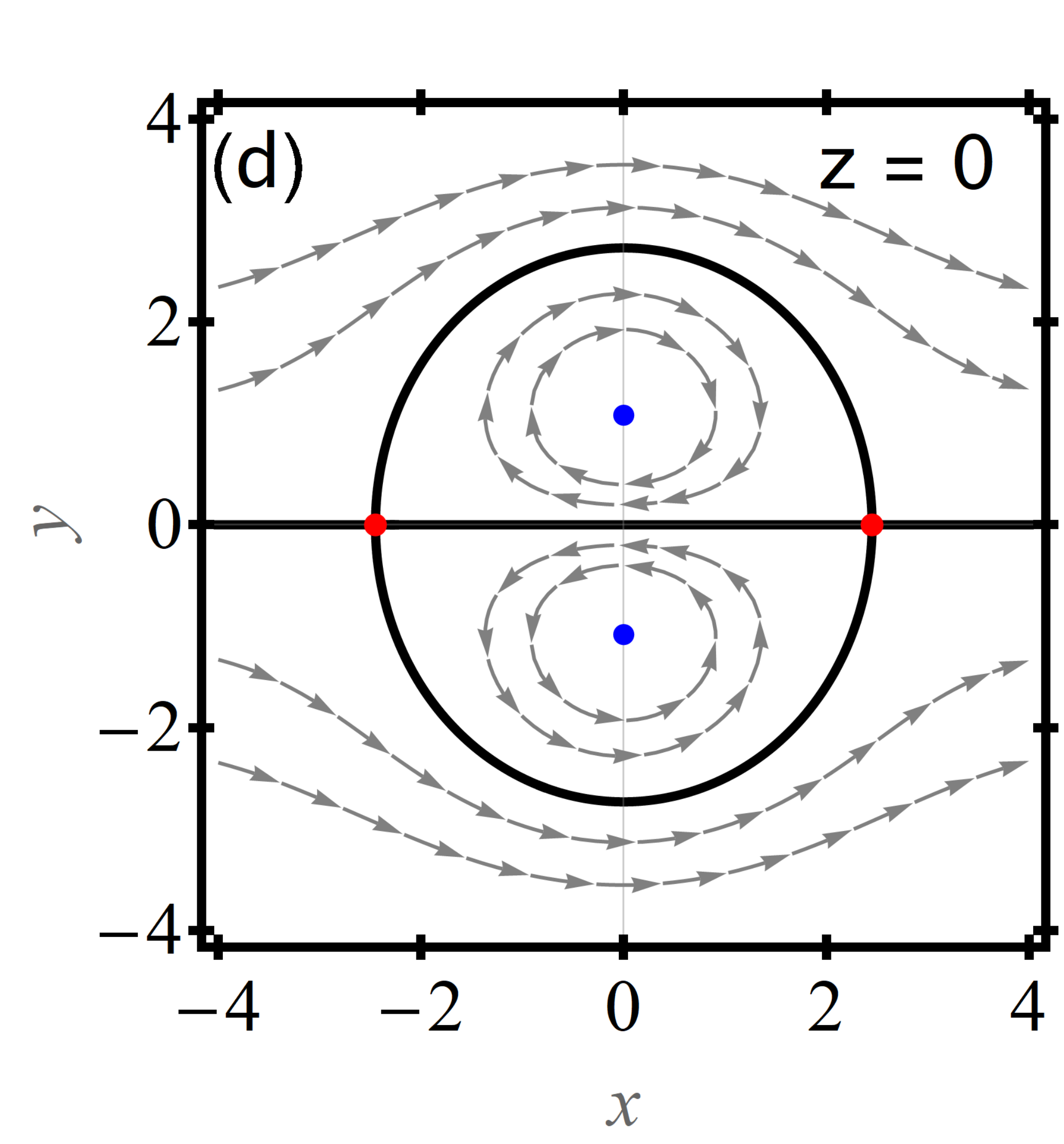}
\caption{Phase portraits of the flow induced by a heton with $Z = 1$ on horizontal planes with (a) $z = -4$, (b) $z = -2$, (c) $z = -1$, and (d) $z = 0$. The legend remains the same as in Figure \ref{fig2}, except that the separatrix is given by $\psi = \psi_u$, where $\psi_u$ is the value of the streamfunction at an unstable fixed point. No fluid is trapped sufficiently far away from the heton (a), but a finite trapping region bounded by a homoclinic separatrix is seen to exist upon increasing $z$ (b). Further increase in $z$ shows the presence of a heteroclinically bound trapping region (c). Finally, at $z = 0$, the trapping region is seen to be symmetric about the $x$ and $y$ axes, like in the $Z = 0$ case (d).}
\label{fig4}
\end{figure}

The phase portraits now show three different structures (Figure \ref{fig4}). Again, far above and below the heton, $z\ll -Z$, and $z \gg Z$, the heton induced velocity is small, the flow is dominated by the drift, there are no fixed points, and no trapping (Figure \ref{fig4}a).

\FloatBarrier  

Again, treating $z$ as a bifurcation parameter and moving up from $z=-\infty$, we start with no fixed points. We then reach a range of $z$ with two fixed points, one stable and one unstable, with a trapping region bounded by a homoclinic connection (Figure \ref{fig4}b). As $z$ moves closer to zero we find a range of heights with a structure similar to that previously seen when $Z=0$. This region has two stable centers, two unstable saddles, and a trapping region bounded by heteroclinic manifolds. Unlike the $Z=0$ case, the phase portrait is only symmetric in $y$ when $z=0$. For each height where a trapping region exists, the trapping region is bounded by the contour $\psi = \psi_{u}(z)$, where the subscript denotes the streamfunction value at the unstable fixed point at the height $z$. 

\begin{figure}[H]
\centering
\includegraphics[width=0.88\linewidth]{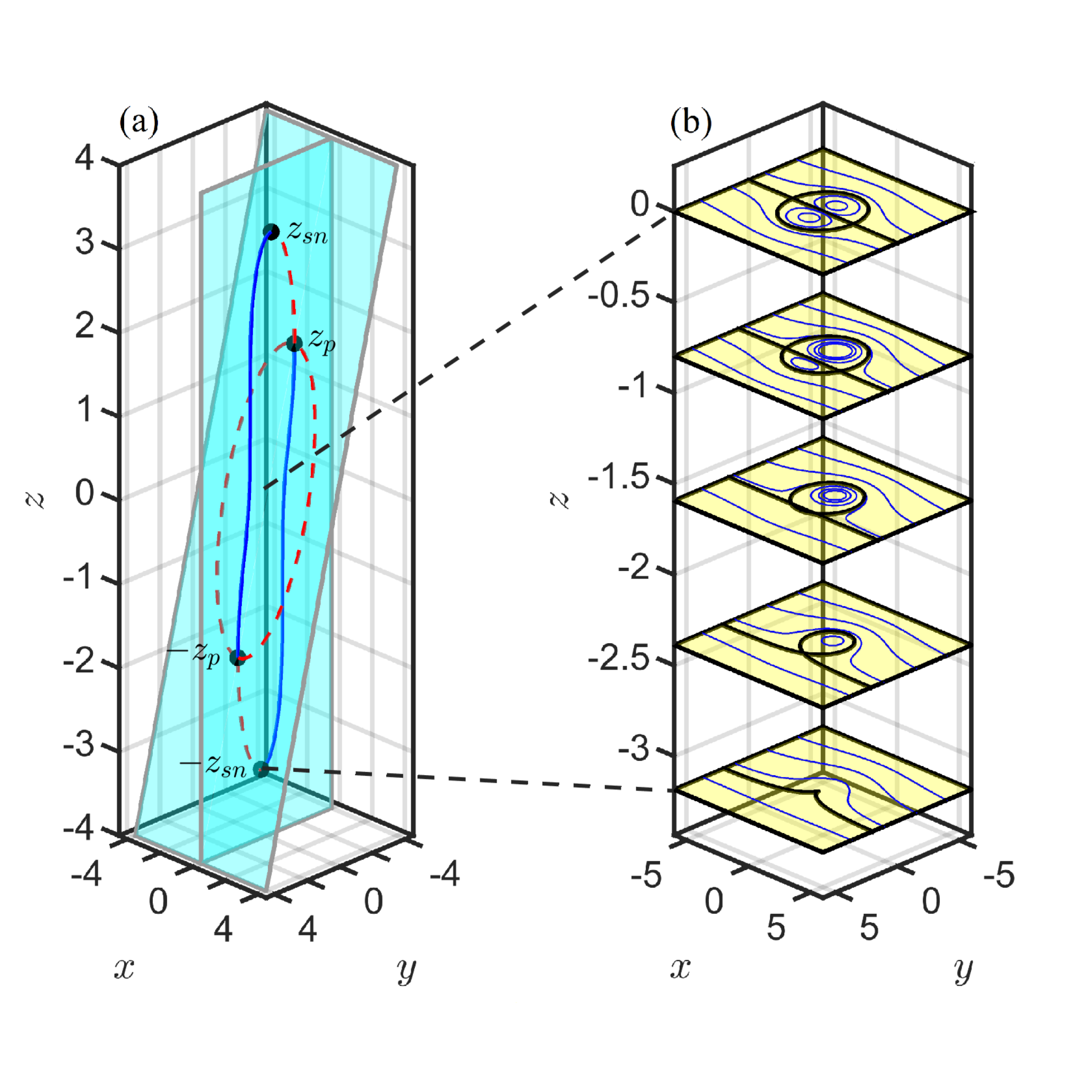}
\caption{Bifurcation diagram (a) and contour slices (b) for the $Z = 1$ case. Panel (a) illustrates the generation/destruction of stable (solid blue curves) and unstable (dashed red curves) fixed points. At $z = -z_{sn} \approx -3.18$, a saddle-node bifurcation leads to the creation of a saddle and a center, both on the plane $x = 0$. At $z = -z_p = - \sqrt{3}$, a pitchfork bifurcation causes the saddle to give way to a center on $x = 0$ and two saddles on $y = -z$. Similar bifurcations occur for $z > 0$, destroying the fixed points. The contour slices of panel (b) illustrate the three-dimensional volume trapped by the heton at five equally spaced planes between $z = -3.18$ and $z = 0$. 
}
\label{fig5}
\end{figure}

\begin{figure}[H]
\centering
\includegraphics[width=0.88\linewidth]{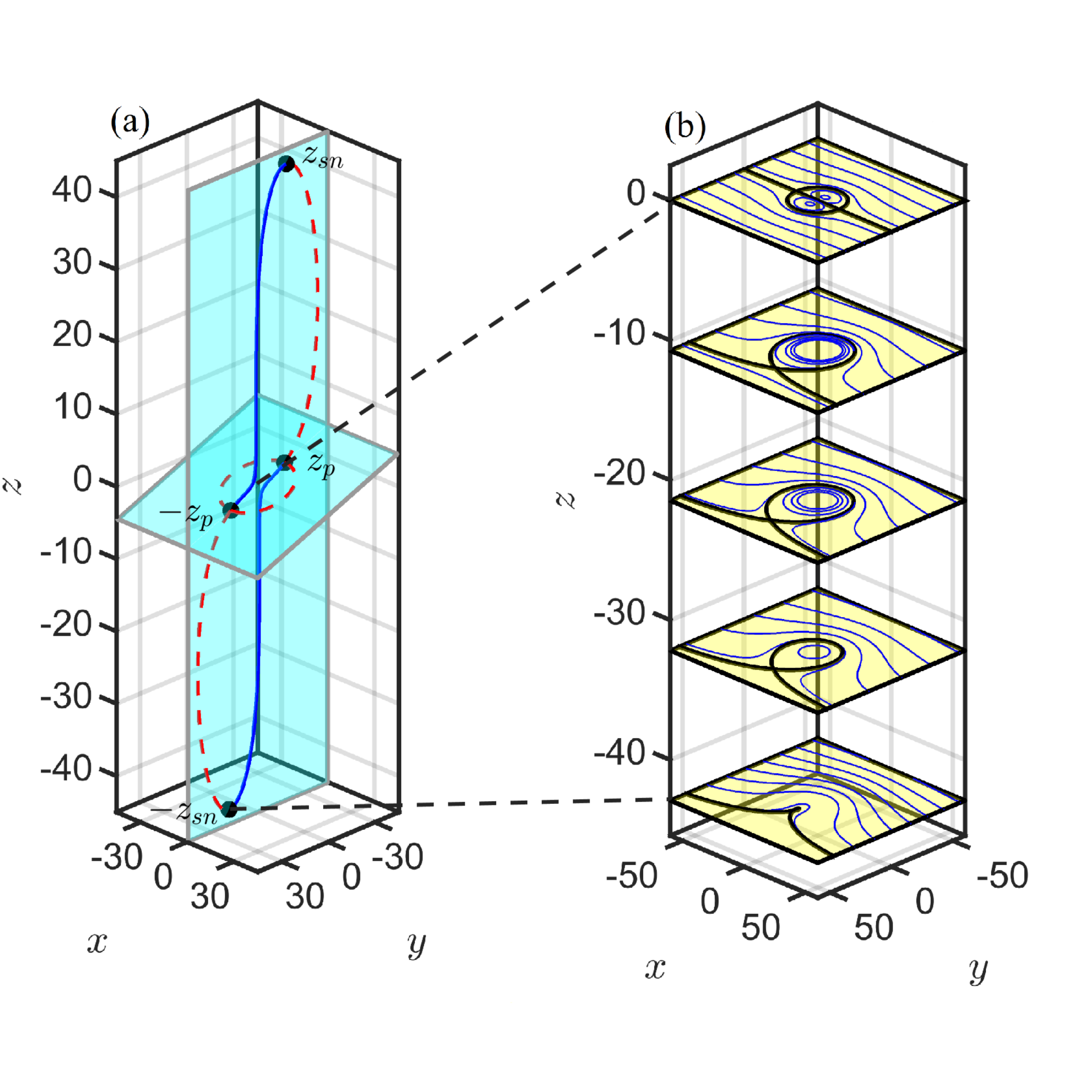}
\caption{Same as Figure \ref{fig5} but for $Z=10$. The saddle-node bifurcation is now at $z_{sn} \approx 42.96$. The pitchfork bifurcation remains at $z_p =  \sqrt{3}$ independent of $Z$.}
\label{fig6}
\end{figure}

The bifurcation diagrams for $Z\ne0$ (Figures \ref{fig5}a, \ref{fig6}a) show the bifurcation sequence. As one moves up from $z=-\infty$, there is saddle-node bifurcation at $z=-z_{sn}(Z)$, which is now a function of $Z$, creating a single stable center and a single unstable saddle with a homoclinic connection bounding the trapping region. As one moves up from $z=-z_{sn}$, there is a second bifurcation, a pitchfork bifurcation, at $z=-z_p$, where the unstable fixed point changes stability to become a stable center and creates two new unstable saddles with a heteroclinic connection. The centers invariably lie on the $y$-axis, as in the $Z=0$ case. The saddles, which had $y = 0$ for all $z$ when $Z=0$, now have $y\ne 0$ for $z \ne 0$. Increasing $z$ through zero to $+\infty$ reverses the bifurcation sequence with a subcritical pitchfork bifurcation at $z=+z_p$ and a saddle-node bifurcation at $z=+z_{sn}$. 
\FloatBarrier
\FloatBarrier

The 3D fluid surrounding each tilted heton thus has three distinct dynamical regimes depending on the height $z$ in the fluid relative to the heton center plane:
\begin{enumerate}
    \item A \emph{heteroclinic regime}, $|z| < z_{p} = \sqrt{3}$. This regime is characterized by the presence of four fixed points, two stable, and two unstable. The unstable fixed points are connected by heteroclinic orbits which bound the trapping region in each horizontal plane.
    
    \item A \emph{homoclinic regime}, $ z_{p} \leq |z| \leq z_{sn}(Z)$. This regime is characterized by the presence of two fixed points, one stable, and one unstable. The unstable fixed point has an associated homoclinic orbit which bounds the trapping region in each horizontal plane.
    \item A \emph{drift regime}, $|z| > z_{sn}(Z)$. This regime has no fixed points and no trapping. 
\end{enumerate}
These regions are illustrated in Figures \ref{fig4}, \ref{fig5}b and \ref{fig6}b.

The unstable fixed points in the heteroclinic regime can be found analytically by setting the expressions (\ref{hetmovxII}, \ref{hetmovyII}) to zero for nonzero values of $x$. This produces the roots
\begin{align}
    \x = \left(\pm \sqrt{(3 - z^2)(1 + Z^2)},-zZ, z\right).\label{versnfp}
\end{align}
These roots are real when $|z| < \sqrt{3}$ and thus the pithfork bifurcation occurs at $z_{p} = \sqrt{3}$ independent of $Z$. We note that this is the same height as the saddle-node bifurcation when $Z=0$, i.e., $z_p = z_{sn}(0)$. The other fixed points are not amenable to analytical discovery, and must be computed numerically (see Figures \ref{fig5}a, \ref{fig6}a).

\section{Volume trapped by a tilted heton, $Z \ne 0$}
The fixed points and manifolds in tilted heton, and therefore the volume $V_t$ of its trapping region, depend on the heton's vertical half-separation $Z$. The contour slices of Figures \ref{fig5}b and \ref{fig6}b roughly illustrate the shape of the 3D trapped volume as well as its growth size as $Z$ increases. Here, we propose a scaling theory for the trapped volume $V_t(Z)$ for $Z >> 1$ and compare with numerical computations.

\subsection{Scaling Theory for $Z\gg1$}
The trapped volume is defined by the interplay between  two parts of the streamfunction (\ref{hetmstrII}). The first two terms in (\ref{hetmstrII}) are contributions from the component vortices of the heton, $ \psi_\pm = 1/4\pi d_\pm$,
where $d_\pm = \sqrt{x^2 - (y\pm 1)^2 + (z \pm Z)^2}$ is the distance from the position $\mathbf{x}$ of a passive particle to the vortex with circulation $\pm 1$. These vortex contributions are denoted $\psi_v = \psi_- - \psi_+$. The third term is a drift term due to being in a co-moving frame,
\begin{equation}
    \psi_d = - \frac{y}{16\pi(1 + Z^2)^{3/2}}.
\end{equation}
Writing the streamfunction as $\psi = \psi_v + \psi_d$, we build a scaling theory for $V_t$ from three assumptions:
\begin{enumerate}
    \item The scaling of the streamfunction determines the scaling of the trapped volume;
    \item The vortex and drift components of the streamfunction scale uniformly, $\psi_v \sim \psi_d$;
    \item The scaling is isotropic, $\mathbf{x} \sim Z^\alpha$ with $\alpha > 0$.
\end{enumerate}
The first assumption is motivated by the fact that the manifolds of the streamfunction determine the trapped volume. The trapped volume should thus scale with the streamfunction. The second assumption is based on the idea that the boundary of the trapped volume is the crossover region where drift and vortex induced velocities balance each other. We thus expect $\psi_v \sim \psi_d$ for the streamfunction values on the manifolds. The final assumption, isotropic scaling, is a reasonable simplification, although, as QG dynamics is anisotropic between the horizontal and vertical, anisotropic scaling would not be unphysical. 

We define nondimensional scaled coordinates $\mathbf{x'} = \mathbf{x}/Z^\alpha$, and then drop the primes. The drift term in scaled coordinates is
\begin{align*}
    \psi_d &= - \frac{y Z^\alpha}{16\pi(1 + Z^2)^{3/2}},\\
    &\sim Z^{\alpha-3}.
\end{align*}
The squared-distance from the vortex is 
\begin{equation}
    d^2_\pm = r^2 Z^{2\alpha} \pm 2 y Z^{\alpha} \pm 2 z Z^{\alpha+1} + 1 + Z^2,
    \label{dpm}
\end{equation}
where $r^2 = x^2+y^2+z^2$. The largest term in $d^2_\pm$, i.e., the term with the largest exponent on $Z$, depends on the value of $\alpha$. Define the largest term as $d^2_{\pm_0}$. Then
\begin{align}
    d^2_{\pm_0} = 
    \begin{cases}
    Z^2, & \alpha < 1, \\
     (r^2 \pm 2 z + 1) Z^2, &\alpha = 1\\
    r^2 Z^{2\alpha}, & \alpha > 1. 
    \end{cases}
    \label{dpm0}
\end{align}
The total squared distance can be written as $d^2_\pm = d^2_{\pm_0}(1 + \epsilon_\pm)$ where $\epsilon_\pm$ is obtained from Eqs.(\ref{dpm}) and (\ref{dpm0}) by $\epsilon_\pm = (d^2_\pm-d^2_{\pm_0})/d^2_{\pm_0} \ll 1 $.
Expanding $\psi_\pm$ gives
\begin{align}
    \psi_\pm 
    &=\frac{1}{4 \pi d_{\pm_0}}(1 - \epsilon_\pm/2+ O(\epsilon^2_\pm) ),
\end{align}
and then 
\begin{align}
    \psi_v &= \left(\frac{1}{d_{-_0}}- \frac{1}{d_{+_0}}\right) -\frac{1}{2} \left(\epsilon_- - \epsilon_+\right) + \ldots.
\end{align}

For $\alpha = 1$, $d_{+_0} \ne d_{-_0}$ and the leading order contribution to $\psi_v$ is given by $\psi_v\sim 1/d_{-_0} - 1/d_{+_0} \sim Z^{-1}$. In this case, $\alpha = 1$,  $\psi_d\sim Z^{-2}$ and thus the assumption that $\psi_v\sim\psi_d$ cannot be satisfied: $\alpha = 1$ is inconsistent with the scaling theory.

When $\alpha < 1$ or $\alpha >1$, $d_{+_0} = d_{-_0}$, and the leading order term in $\psi_d$ is given by the leading order term in $\epsilon_- - \epsilon_+$,
\begin{equation}
    \psi_v = 
    \begin{cases}
    z Z^{\alpha - 2}/2\pi + \ldots, & \alpha < 1,\\
     z Z^{1-2\alpha}/2\pi r^3 + \ldots, & \alpha > 1
    \end{cases}
\end{equation}
For the case $\alpha < 1$, $\psi_v\sim \psi_d$ requires $\alpha - 2 = \alpha-3$ which is has no solutions. Thus, $\alpha<1$ is also inconsistent with the theory. The final case, $\alpha > 1$ leads to $1-2\alpha = \alpha - 3$ which has one solution $\alpha = 4/3$, and is consistent with the requirement $\alpha > 1$ for the case being considered. Thus, isotropic scaling theory leads to a single scaling exponent, $\alpha = 4/3$. Then $\mathbf{x}\sim Z^{4/3}$ and $V_t \sim x\, y\, z \sim Z^4$. 
The dimensional trapped volume then scales $Z^4/Y$.

\subsection{Numerical calculation of the trapped volume}
\begin{figure}[H]
\includegraphics[width=0.5\linewidth]{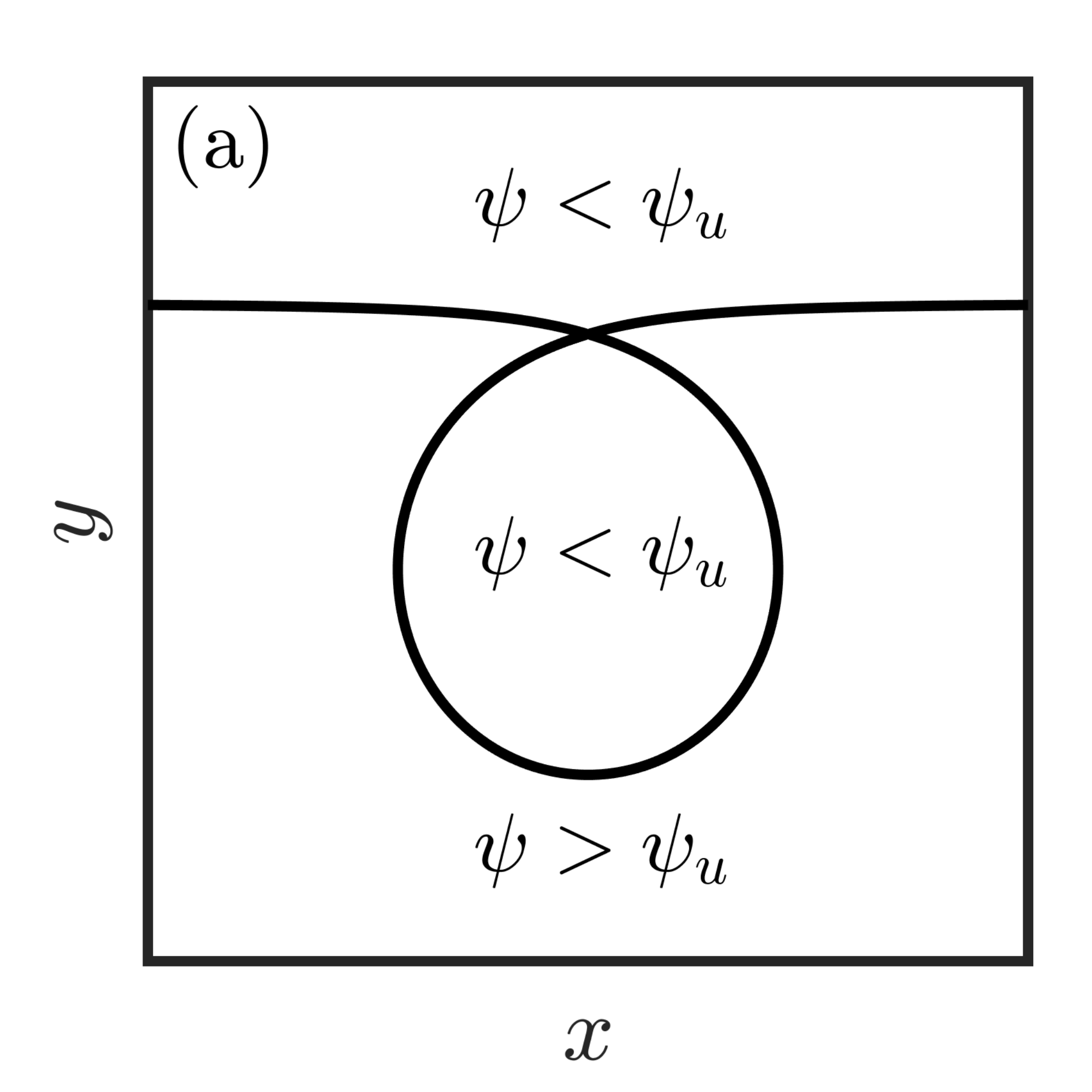}
\includegraphics[width=0.5\linewidth]{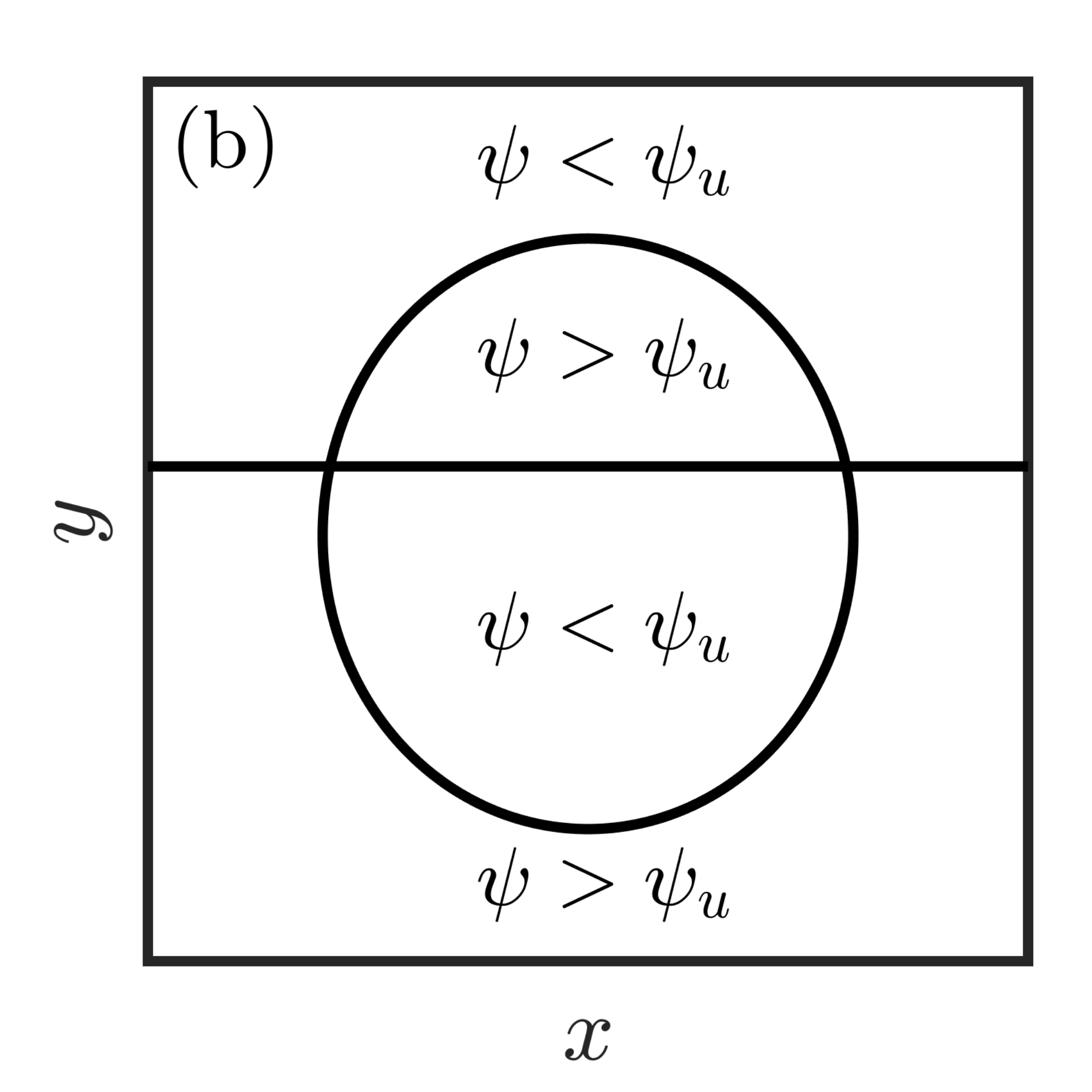}
\caption{Schematic illustrating the geometry of the streamfunction values and the structure of the invariant manifolds at a fixed height $z$ defining the trapped volume in the (a) homoclinic and (b) heteroclinic regime for $z < 0$. For $z > 0$, the structures are flipped across $y = 0$, and regions inside heteroclinic separatrices display the same pattern of streamfunction values, but regions inside homoclinic separatrices have $\psi > \psi_u$. Here $\psi_u$ is the streamfunction value at an unstable fixed point. Note that when $Z = 0$, there is no homoclinic regime, the trapped volume is symmetric about all three coordinate planes, and $\psi_u = 0$.}
\label{fig7}
\end{figure}

We have already seen that for vertically aligned hetons, $Y=0$, the trapped volume is, in some sense, infinite. The trapped volume in the remaining two cases, defined by $Y=1$ and a single value of $Z$, can be calculated numerically. The trapped region is bounded by the invariant manifolds of the unstable fixed points. These manifolds are streamlines with streamfunction $\psi = \psi_u(z)$ where $\psi_u(z)$ is the value of the streamfunction at the associated unstable fixed point at height $z$. The values of $\psi$ relative to $\psi_u$ in the homoclinic and heteroclinic regimes are shown in Figure \ref{fig7}.
\FloatBarrier
To calculate the trapped volume $V_t(Z)$, we numerically solve (\ref{hetmovxII}-\ref{hetmovyII}) for the unstable fixed points over a range of $z$ from $z = 0$, where we know the fixed points exist, to $z = z_{sn}$, where the fixed points have vanished and no fluid is trapped. At each value of $z$ we calculate $\psi_u$ from (\ref{hetmstrII}) and demarcate the trapped region using the geometry in Figure \ref{fig7}. The trapped area bounded by $\psi = \psi_u(z)$ is then found by numerical integration. Further numerical integration of the trapped area over the heights where trapped fluid exists gives the trapped volume. Solving for the fixed points requires balancing two numerically calculated terms. As $Z$ becomes large, these terms become large and numerical error limits the accuracy of the solution. As a result we restrict our calculation of the trapped volume to the range $0 \le Z \le 10$.

\begin{figure}[H]
\includegraphics[width=0.5\linewidth]{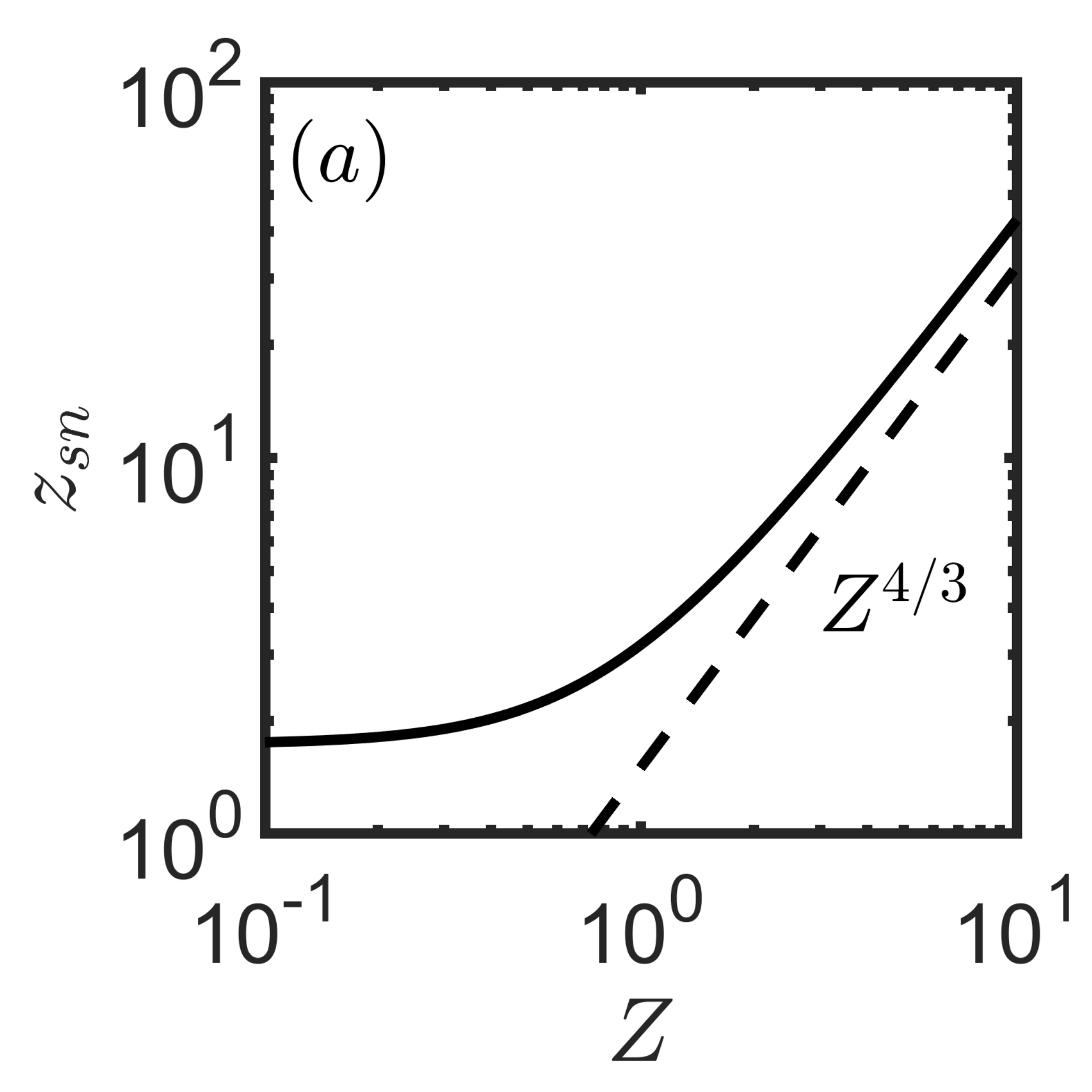}
\includegraphics[width=0.5\linewidth]{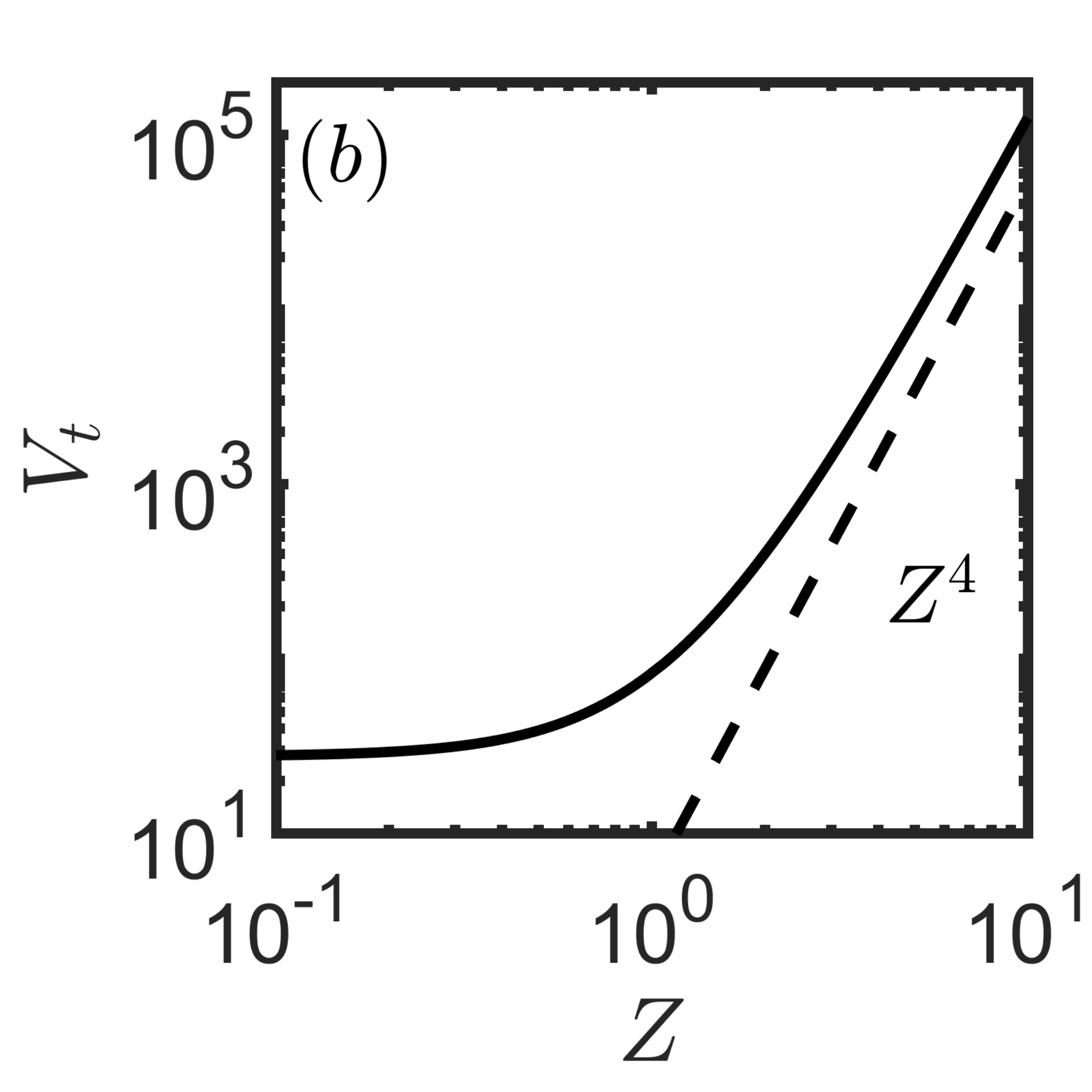}
\caption{Logarithmic plots showing the variation of the (a) saddle-node bifurcation point $z_{sn}$ and (b) trapped volume $V_T$ with $Z$ (solid line). Both plots conform to the expected scaling (dashed line).}
\label{fig8}
\end{figure}

The height of the saddle-node bifurcation $z_{sn}$ and the trapped volume $V_t$ are shown in Figure \ref{fig8}. We see that the scaling theory is approximately verified, with the growth slightly shallower than the scaling theory predicts. We conjecture that at larger $Z$ the scaling theory becomes more accurate.

\FloatBarrier

\section{Discussion}

Since their inception in the papers by Gryanik \cite{gryanik_dynamics_1983-1, gryanik_dynamics_1983}, and Hogg and Stommel \cite{hogg_heton_1985}, point vortex hetons have been used to model a variety of geophysical transport phenomena. Central to these studies is the fact that a moving heton carries with it a finite volume of fluid. In this paper, we have characterized the volume trapped by a single 3D heton in an infinite, continuously stratified fluid with no background flow, and constant Coriolis and Brunt-V\"{a}is\"{a}l\"{a} frequencies.

A heton is fully defined by the 3D position of each vortex and their common circulation. Translational symmetry in space and time, horizontal isotropy, and the choice of a lengthscale allow all hetons to be characterized by vortex separation alone. The 3D fluid dynamics of each heton configuration foliates into a family of 2D horizontal Hamiltonian systems parametrically dependent on the heton configuration parameter and the vertical distance in the fluid from the center of the heton. 

The possible heton configurations separate into three categories with distinct transport properties. \emph{Vertically aligned hetons} have zero horizontal separations and are all similar as the vertical separation can be scaled to unity. They are stationary and induce axisymmetric rotation about the $z$-axis. Since there is no translation of fluid relative to the heton there is no long-range transport and we might say that the trapped volume is infinite. The remaining two configurations have well-defined three dimensional trapping regions.  \emph{Horizontally aligned hetons} have vortices with zero vertical separation and are also all similar as the horizontal half-separation $Y$ can be scaled to unity. Horizontally aligned hetons exhibit trapping in the region of the fluid where the dimensional height $z$ satisfies $|z| < \sqrt{3}Y$. The trapping region is bound in each horizontal plane by a heteroclinic separatrix symmetric across the line bisecting the vortex separation. The more general \emph{tilted hetons}, hetons whose vortices have nonzero horizontal and vertical separations, are characterized by a single parameter, the ratio of the vertical half-separation to the horizontal half-separation, $Z/Y$. These hetons exhibit a trapping region that is bounded by heteroclinic separatrices for regions of the fluid with dimensional height $|z| < \sqrt{3} Y$, where a pitchfork bifucation occurs, and by homoclinic separatrices when $\sqrt{3}Y \leq z \leq z_{sn}Y$, where $z_{sn}$ is the height were a saddle-node bifurcation occurs. From scaling theory and numerical simulations, we find the dimensional trapped volume scales as $Z^4/Y$ for $Z >> 1$.

Two common idealizations of fluid stratification are the 3D uniform stratification studied here and a two-layer QG fluid. Transport by two-layer hetons shows some analogs with the results found here \cite{hogg_heton_1985,young_interactions_1985}. Two layer hetons are characterized by a single parameter, the ratio between the horizontal vortex separation and the interfacial Rossby radius. Young \cite{young_interactions_1985} showed that when the vortices are both in the upper layer, there is always trapped volume in the upper layer and a trapped volume in the lower layer only when the separation parameter is below a threshold. This is analagous to our horizontally aligned heton where the trapped volume extends a finite height above and below the plane containing the heton. The two-layer analogue of our tilted heton is a heton with one vortex in each layer. Young \cite{young_interactions_1985} showed that every such heton is in either a homoclinic or heteroclinic regime depending on the horizontal vortex separation. This single regime fills the entire fluid with streamfunctions in the two layers being the opposite of each other, $\psi_1 = -\psi_2$. This differs from the 3D heton studied here in that we find every tilted heton configuration has all three regions described above. In the two-layer tilted heton analog, one can never experience a height without a vortex, and hence the entire fluid displays trapping. In the 3D case, the trapped volume grows without bound as the vertical vortex separation grows with fixed horizontal separation. This behavior has an analogue in the two-layer case where the trapped volume grows without bound as the horizontal separation decreases  \cite{hogg_heton_1985, young_interactions_1985}. 

The trapped region in the QG heton is trapped forever and the region is defined by the heteroclinic/homoclinic connections in the phase space of the Hamiltonian. If the Hamiltonian is perturbed it becomes possible for these connections to break, leading to Lagrangian chaos and a region with finite time trapping and L\'evy flights \cite{weiss_mass_1989, solomon_observation_1993}. One physically interesting perturbation is the inclusion of ageostrophic effects, which bring in  small, order Rossby number, corrections, including a small, nonzero, vertical velocity. A recently developed theory for ageostrophic 3D point vortices \cite{weiss_point_2022} will allow the study of ageostrophic perturbations to heton and may show finite time trapping and L\'evy flights. Other interesting perturbations include a fluid with more point vortices than a single heton, and generalizing the point vortex heton to hetons with distributed vortices.

\acknowledgments{The authors would like to thank James Meiss and Ian Grooms for helpful discussions.}







\reftitle{References}
\externalbibliography{yes}
\bibliography{Definitions/template.bib}


\end{paracol}

\end{document}